\documentclass[%
 reprint,
 amsmath,
 amssymb,
 aps,
 pra,
 floatfix,
]{revtex4-2}
\usepackage{url}
\usepackage{graphicx}
\usepackage{algorithm}
\usepackage{algpseudocode}
\usepackage{braket}
\usepackage{dcolumn} 
\usepackage{bm}

\begin{document}

\preprint{APS/123-QED}

\title{Statistical Qubit Freezing Extending Physical Limit of Quantum Annealers}



\author{Jeung Rac Lee}
\author{June-Koo Kevin Rhee}
\email{rhee.jk@qunovacomputing.com}
\affiliation{%
 Qunova Computing, Inc., Daejeon, 34130, South Korea \\
 KAIST, Daejeon, 34141, South Korea
}%

\author{Changjun Kim}
\author{Bo Hyun Choi}
\affiliation{%
 LG Uplus, Seoul, 07795, South Korea
}

\date{\today}

\begin{abstract}

Adiabatic quantum annealers encounter scalability challenges due to exponentially fast diminishing energy gaps between ground and excited states with qubit-count increase. This introduces errors in identifying ground states compounded by a thermal noise. We propose a novel algorithmic scheme called statistical qubit freezing (SQF) that selectively fixes the state of statistically deterministic qubit in the annealing Hamiltonian model of the given problem. Applying freezing repeatedly, SQF significantly enhances the spectral gap between of an adiabatic process, as an example, by up to 60\% compared to traditional annealing methods in the standard D-Wave's quantum Ising machine solution, effectively overcoming the fundamental limitations.
\end{abstract}

\maketitle


\section{\label{sec:intro} Introduction}

Quantum annealing (QA) can leverage quantum tunneling, potentially to realize a quantum advantage in determining the ground state energy of a system, compared to traditional classical annealing methods ~\cite{kadowaki1998, johnson2011, jiang2022}. It represents an emerging paradigm in tackling optimization and sampling problems challenged by classical computers. Unlike universal gate-based quantum computers, quantum annealers utilize adiabatic quantum computation, transitioning from an initial Hamiltonian to a target Hamiltonian to solve given problems. While manifesting significant potential in diverse applications such as combinatorial optimization, machine learning, and material science~\cite{karimi2012,amin2018,kitai2020}, their practical applications often face scalability limits due to environmental noises, limited qubit counts, and coherence times~\cite{albash2018, stilck2021}.

The adiabatic theorem states that the evolution time to guarantee the ground state solution is inversely proportional to a spectral gap of the Hamiltonian~\cite{born1928}. However, as the problem size increases, the spectral gap of the system diminishes exponentially~\cite{albash2018,mishra2018}. Consequently, scalability limit arises when a quantum advantage is pursued using naive adiabatic quantum computation in a non-zero temperature limit, both theoretically and empirically~\cite{albash1983, stilck2021}. The limited coherence time of a typical Ising machine further prevents QA processes from meeting the adiabatic criteria for sufficiently extended evolution times; thermal noises exacerbate the challenge of maintaining the system in the ground state. Additionally, a common realization of a quantum adiabatic process lacks qubit error correction capabilities. 

In order to circumvent these limitations, various approaches have been explored. As an example, encoding logical qubits across numerous physical qubits can mitigate environmental noises and control errors.~\cite{vinci2016, vinci2018}. Even though this strategy has shown theoretical and empirical improvements to some extent, its scalable applications remain questionable because it requires a cluster of multiple qubits to form a logical qubit, which is a critical limit to achieve quantum advantage with a noisy intermediate-scale quantum (NISQ) devices~\cite{preskill2018}. 


In this paper, we investigate an iterative scheme called Statistical Qubit Freezing (SQF), which aims to gradually widen the spectral energy gap by excluding qubits from the original problem when their values for the solution become obvious with high probability. Early-stage investigation of SQF was suggested in previous work~\cite{jeung2023}, and a similar methodology called Iterative Sample Persistence Variable Reduction (ISPVAR) has been introduced in other studies ~\cite{karimi2017_1, karimi2017_2}. In this SQF scheme, the number of deterministically frozen qubits increases iteratively, thereby reducing the size of the QA problem. This iterative process enables finding optimal or near-optimal solutions, even for problems surpassing the physical limits of a quantum annealing process. Compared to previous methodologies, the SQF scheme presented in this paper is distinct due to its verification steps, which assesses energy differences of freezing variables before excluding them and considers the coupling strength and statistical properties of neighboring variables. 

In a typical embodiment of a QA system, the full scale of the system energy remains finite and constant regardless of the number of active qubits in an adiabatic process. Hence, the corresponding energy gaps diminish exponentially as the number of qubits increases. Conversely, energy gaps are expected to widen if a portion of qubits are removed from the original QA problem. In an SQF scheme, a qubit is eliminated from the QA problem when sampling of QA results suggests the qubit can be classically fixed to a boolean value. This freezing process can be repeated to increase the number of frozen qubits until the QA process reaches the lowest energy solution.

\section{Ising Hamiltonian of the system}

We validate the performance of the proposed algorithm using a commercially available quantum annealer provided by D-Wave Systems. 
D-Wave offers a cloud service granting access to the D-Wave Advantage, equipped with over 5,000 qubits and more than 35,000 couplers~\cite{dwaveleap, dwaveadvantage}. This quantum processor is manufactured in a specific configuration named \textit{Pegasus}, where each qubit is connected to 15 other qubits~\cite{dwavepegasus}. 

\begin{figure}[ht]
\includegraphics[width=0.5\textwidth]
{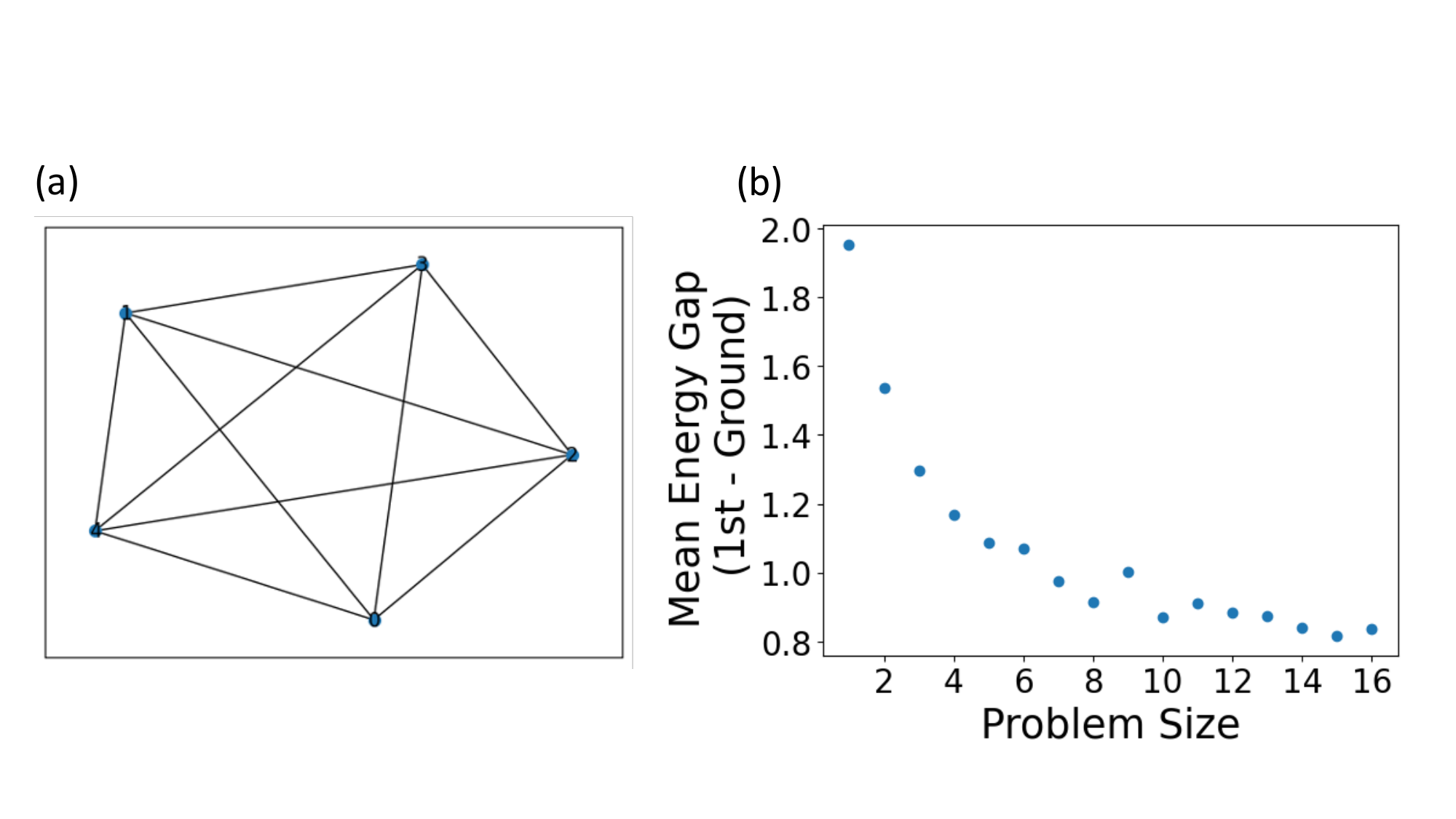}
\caption{\label{fig:system} Schematics of the system behavior. (a) A simple example of a five-qubit random Ising system with complete connectivity. (b) energy gap between the first excited and ground states versus qubit counts that shows an exponential decrease of the energy gap.}
\end{figure}

To solve an optimization problem via a D-Wave machine, the problem of interest first needs to be formulated as an Ising problem, or as a Quadratic Unconstrained Binary Optimization(QUBO) problem. The two problems are mathematically equivalent through the linear transformation of the variables, $s_i = 2x_i -1$, where $s_i \in \{-1, 1\}$ and $x_i \in \{0, 1\}$. 
For simplicity, we will only consider Ising problems for the rest of this article. After successfully formulating the problem, due to the limited connectivity of QA devices, the problem then needs to be \textit{minor embedded} to conform to the quantum Ising machine configuration on the quantum process unit (QPU), a process which can be facilitated by the Python library provided by D-Wave Systems~\cite{cai2014}.

Once the problem is successfully embedded on the QPU, the system evolves from an initial state to a final state programmed by the user that encodes the solution as its ground state through a predefined annealing schedule. 
The generalized Ising Hamiltonian with $n$ qubits comprises terms for the initial Hamiltonian $H_\mathrm{I}$ and the final Hamiltonian $H_\mathrm{F}$ \cite{johnson2011}, represented as
\begin{align}
H(s) &= A(s)H_\mathrm{I}+B(s)H_\mathrm{F}, \\
\label{eq:H_ising}
H_\mathrm{I} &= -\sum_{i}\hat{\sigma}_{x}^{(i)}, \\
H_\mathrm{F} &= \sum_{i}h_{i}\hat{\sigma}_{z}^{(i)}+\sum_{i>j}J_{i,j}\hat{\sigma}_{z}^{(i)}\hat{\sigma}_{z}^{(j)},
\end{align}
where, $\hat{\sigma}_{x}^{(i)}$ denotes the Pauli-$X$ operator acting on the $i$-th qubit $\ket{\psi_i}$, $\hat{\sigma}_{z}^{(i)}$ the Pauli-$Z$ operator, $h_{i}$ the bias, and $J_{i, j}$ the coupling strength between the $i$-th and $j$-th qubits. Each qubit can be represented by $\ket{\psi_i}\equiv a\ket{0}_i+b\ket{1}_i$, where $|a|^2+|b|^2=1$.  Here the quantum state is denoted by $\ket{\Psi}=\bigotimes_{i=1..n}\ket{\psi_i}\equiv \ket{\psi_1\psi_2...\psi_n}$. 

The non-negative real functions $A(s)$ and $B(s)$ define the predefined annealing schedule, where $s = t/t_\mathrm{anneal}$, $0\leq t \leq t_\mathrm{anneal}$ denotes the normalized evolution time coordinate, or {\em anneal fraction}.

In the annealing process, the initial Hamiltonian dominates by a large $A(s)$ value, creating the corresponding ground state $\ket{\Psi_0}=\ket{+}\equiv\bigotimes_{i=1..n}\ket{+}_i$, where $\ket{+}_i = 1/\sqrt{2} (\ket{0}_i+\ket{1}_i)$, which is a superposition of all possible states with respect to the $Z$ basis.The corresponding expectation of the Hamiltonian in the initial state is given $\braket{H(0)}= \braket{\Psi_0|H_\mathrm{I}|\Psi_0}=-nA(0)$. 

 Subsequently, the strength of the initial Hamiltonian governed by $A(s)$ gradually decreases while the strength of a user Hamiltonian governed by $B(s)$ increases, facilitating the gradual evolution of the total Hamiltonian towards the final Hamiltonian given by the user.


\section{Statistical Qubit Freezing (SQF)}

Statistical Qubit Freezing (SQF) aims to improve the solution accuracy of a QA process by fixing, i.e. freezing, initial values of logical variables with a certain level of confidence, thereby reducing the problem size. Freezing of logical variables increases with iterative applications of QA with updated freezing at each step. 
The statistical nature of quantum computation, and inherent quantum errors of a device necessitate running the same calculation thousands of times to obtain highly accurate results. In QA, from a set of samples of results from calculation runs, the one exhibiting the lowest energy is selected as the candidate solution, while the others are disregarded. 
SQF, however, tracks the strings of classical bit values to which qubit converged across all samples of the annealing process results. Consequently, even without knowing the optimal solution, certain qubits tend to settle with high probability into either -1 or 1, allowing users to assume the optimal values with some confidence.

The efficacy of SQF lies in its reduction of the problem size during consecutive annealing cycles. One of the advantages of quantum computations is the exponential increase in the space that such quantum states can represent as the number of qubits grows. 
However, the limited energy scale of the D-Wave quantum annealer device leads to a densely distributed eigenenergy spectrum as the number of eigenstates increases, which corresponds to the size of the tensor product states of qubits. According to the adiabatic theorem, the required evolution time, which is inversely proportional to the spectral gap, becomes unrealistically long to reach the global minimum. 
In addition, narrow energy gaps make the system susceptible to being excited by the interaction with an environment. SQF mitigates this behavior by initially avoiding superposed eigenstates unlikely to be solution states, stretching gaps in the energy spectrum, and bolstering the system against undesirable excitation.

We conducted experiments applying SQF to two different problem types. The first comprises random Ising instances where each variable connects to all others in a complete graph as shown in Fig.~\ref{fig:system}(a) illustrating an example with five qubits. 
The second problem type is the Not-All-Equal 3 Satisfiability (NAE3SAT) problem. D-Wave Systems Ocean SDK incorporates an NAE3SAT implementation directly compatible with the quantum annealer~\cite{dwavenae3sat}. This NAE3SAT implementation is designed in a way that the energy of the ground state can be determined in advance so that it is convenient to identify how close the computed results are to the ground state. 

\renewcommand{\algorithmicrequire}{\textbf{Input:}}
\renewcommand{\algorithmicensure}{\textbf{Output:}}

\begin{algorithm}[H]
\caption{Statistical Qubit Freezing (SQF)}\label{alg:SQF}
\begin{algorithmic}[0]
\Require 
    \State {Final Hamiltonian, $H_\mathrm{F}$}
    \State {Freezing Threshold, $z_\mathrm{f}$}
    \State {Sampling Count, $m$}
\Ensure 
    \State {Frozen Hamiltonian, $H^\mathrm{SQF}_\mathrm{F}$}
    \State {Frozen Qubit Set, $F$}
    \State {Frozen Value Set, $Z$}
    \State {Sample Set, $S = \{(s_{k}, e_k)|k = 1.. m\}$}

\State {}
\State {$A \gets H_\mathrm{F}.\mathrm{variables}$}
\State {$F \gets \varnothing$, $Z \gets \varnothing$ }
\Repeat
    \State {$F' \gets \varnothing$}
    \State {$S \gets QA.\mathrm{sample}(H_\mathrm{F}, m)$}
    \ForAll {$i \in A$}
        \State  {Evaluate $z^{(i)}$ from $S_i = \{s_{k,i}|k=1..m\}$.}
    \EndFor
    
    \ForAll {$i\in A$ and $|z^{(i)}| > z_\mathrm{f}$ }
        \State {\bf{if} $z^{(i)} > 0 $ \bf{then} $\bar{z}^{(i)} \gets +1$ \bf{else} $\bar{z}^{(i)} \gets -1$ }
        \ForAll {$j\in A\setminus{\{i\}}$}
            \State {Evaluate~$z^{(j)}$~from}
            \State {$S_j~=~\{s_{k,j}| \forall k~\mathrm{for}~s_{k,i}~=~\bar{z}^{(i)}\}$.}
        \EndFor
        \State {$\delta E_i \gets h_i\bar{z}^{(i)} + \sum_{j \in A \setminus{\{i\}}  } J_{i,j}\bar{z}^{(i)} z^{(j)} $}
        \State {\bf{if} $\delta E_i < 0$ \bf{then} $F' \gets F'\cup \{i\}, ~Z \gets Z \cup \{\bar{z}^{(i)}\} $ }
    \EndFor
    \State {$H_\mathrm{F} \gets H_\mathrm{F}.\mathrm{freeze}(F', Z)$}
    \State {$A \gets A\setminus{F'}$  }
    \State {$F \gets F\cup F'$  }
\Until{$F' == \varnothing$}
\State {$H^\mathrm{SQF}_\mathrm{F} \gets H_\mathrm{F}$} 
\State {\Return {$H^\mathrm{SQF}_\mathrm{F}, ~F,~Z,~S$} }
\end{algorithmic}
\end{algorithm}

\section{Detailed Process of SQF}






The essence of SQF lies in inferring the optimal value of a qubit based on a statistical analysis of quantum annealing outcomes with a certain confidence $\beta$. Let $P_{-1}^{(i)}$ and $P_1^{(i)}$, respectively, denote the probability for the $i$-th qubit to be '-1' or '1', from the statistics sampled from outcomes of a QA. 
Here we define likeliness $z^{(i)} \equiv \braket{\psi_i|\hat{\sigma}_z^{(i)}|\psi_i}= P_{1}^{(i)}-P_{-1}^{(i)}$. In the SQF, qubit $i$ is treated as, or freezed to, '-1' or '1' once $|z^{(i)}|>z_\mathrm{f}$ and freezing helps decrease the energy expectation of the final Hamiltonian. Here, $z_\mathrm{f}$ is called the freezing threshold. Consequently, the count of active qubits decreases as the count of frozen qubits increases, hence enhancing the energy gaps of the corresponding reduced QA. This process can repeat until freezing happens no more.  

\begin{figure*}
\includegraphics[width=0.8\textwidth]
{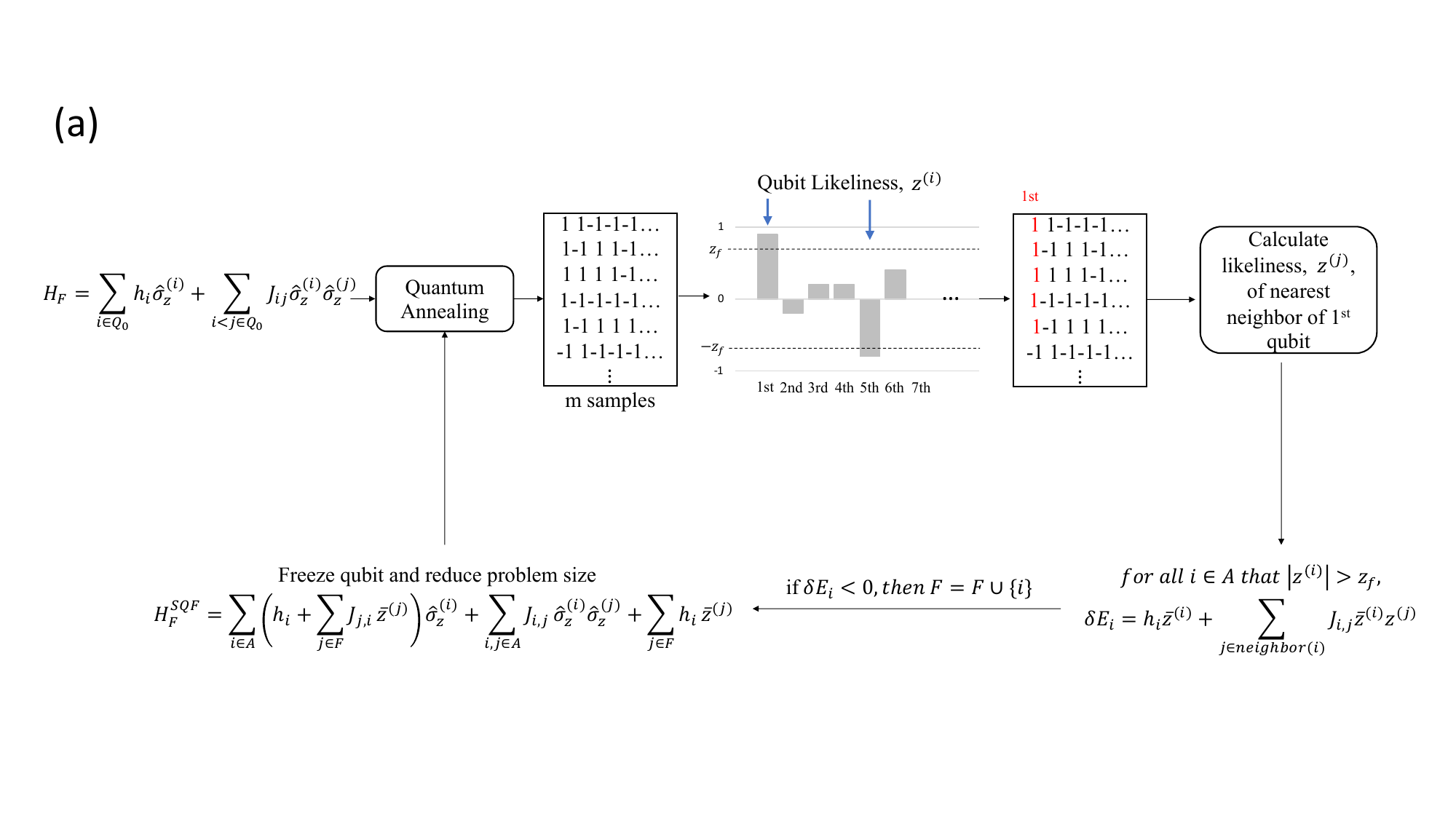}\\
\includegraphics[width=0.8\textwidth]
{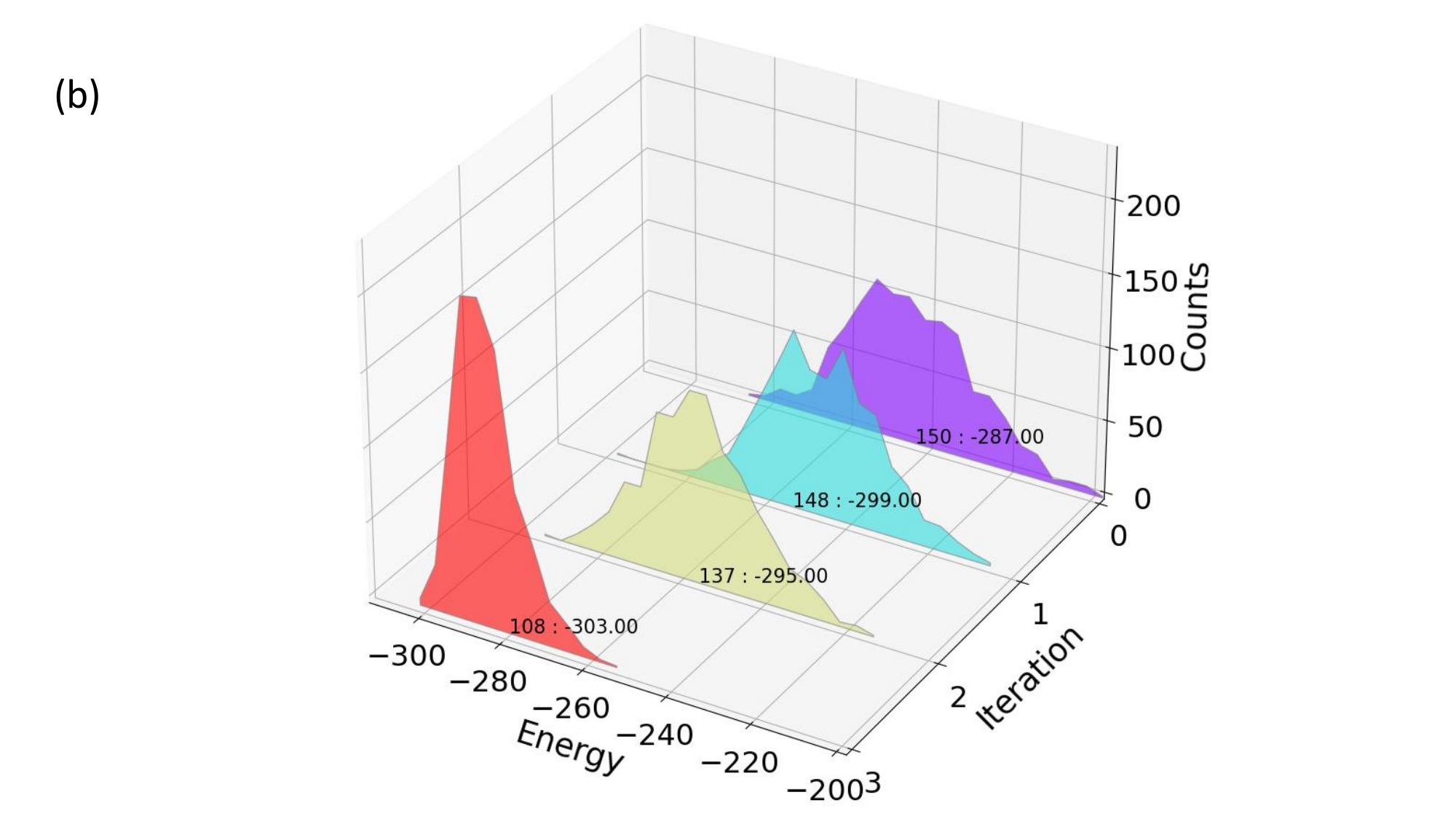}\\
\caption{\label{fig:SQF process} Schematic diagram of the Statistical Qubit Freezing (SQF). (a) Block diagram of the SQF sequence, and (b) the energy histogram after each iteration of SQF. One thousand readouts with the fixed threshold $th=$ 0.6. The results demonstrate the system reaches lower energy states via SQF.}
\end{figure*}

To describe Algorithm 1, consider its application to an Ising problem with a set of $n$ qubits, denoted by indices in the set $Q = {1..n}$. Define $z_f$ be the predetermined freezing threshold. Let $A$ be a set of active qubits obtained by $H_{\mathrm{F}}.\mathrm{variables}$. Let $F$ and $Z$ represent the sets of frozen qubits and their corresponding values, respectively. For each iteration, a QA device samples qubits $m$ times using $QA.\mathrm{sample}$ yielding a sample set $S$ and statistics for each qubit. Then, the likelihood of each qubit, denoted by $z^{(i)}$, is measured. If the absolute value $|z^{(i)}|$ is greater than $z_f$, the qubit is considered for freezing to a classical value $\bar{z}^{(i)}$, determined as $1$ if $z^{(i)}$ is positive and $-1$ if negative. Subsequently, for the qubit $i$, the likelihoods of its neighbors, $z^{(j)}$, are computed from the samples where the value of qubit $i$ matches $\bar{z}^{(i)}$. Once $\bar{z}^{(i)}$ and $z^{(j)}$ are evaluated, compute the freezing merit, $ \delta E_i = h_i\bar{z}^{(i)} + \sum_{j \in A \setminus{\{i\}}  } J_{i,j}\bar{z}^{(i)} z^{(j)}$, to verify if freezing qubit $i$ to $\bar{z}^{(i)}$ is beneficial to the system, meaning if it lowers the energy of the system. If the freezing merit is negative, include qubit $i$ in the set of qubits to freeze, $F'$. After evaluating all qubits, those in $F'$ are frozen using $H_{\mathrm{F}}.\mathrm{freeze}$, setting their states to the respective $\bar{z}^{(i)}$. In the updated Hamiltonian, the biases of the frozen qubits are consolidated as an offset, and their coupling strengths are transferred to the biases of adjacent qubits. The frozen qubits are then removed from $A$ and added to $F$. This process is repeated for the remaining active qubits until no further qubits are left to freeze. Figure~\ref{fig:SQF process}(a) provides a schematic of the SQF process.

After one iteration of SQF, the generalized final Hamiltonian in Eq.~(\ref{eq:H_ising}) becomes 
\begin{widetext}
\begin{subequations}
\begin{align}
H_\mathrm{F}^\mathrm{SQF} &= \sum_{i\in{A}}h_{i}\hat{\sigma}_{z}^{(i)}+\sum_{i,j\in{A}}J_{i,j}\hat{\sigma}_{z}^{(i)}\hat{\sigma}_{z}^{(j)}+\overbrace{\sum_{i\in{F}}h_{i}\bar{z}^{(i)} + \sum_{i\in{F}}\sum_{j\in{A}}J_{i,j}\bar{z}^{(i)}\hat{\sigma}_{z}^{(j)}}^\text{Freezing}
\label{eq:H_ising_SQF} \\
&=\sum_{i\in{A}}(h_{i}+\sum_{j\in{F}}J_{j,i}{\bar{z}}^{(j)})\hat{\sigma}_{z}^{(i)}+\sum_{i,j\in{S}}J_{i,j}\hat{\sigma}_{z}^{(i)}\hat{\sigma}_{z}^{(j)}+\underbrace{\sum_{i\in{F}}h_{i}{\bar{z}}^{(i)}}_\text{offset}.
\label{eq:H_ising_SQF_2}
\end{align}
\end{subequations}
\end{widetext}

Equation (\ref{eq:H_ising_SQF}) represents the finalized SQF Hamiltonian consisting of the frozen sites ("Freezing") and the nearest neighbors. Here, \(A\) represents the set of all active qubits, \(F\) the set of qubits to freeze, \(\bar{z}^{(i)}\) the classical value to which the qubit indexed \(i\) is frozen. The Hamiltonian of frozen qubits is shown in the third term of Eq.~(\ref{eq:H_ising_SQF}), $\sum_{i\in{F}}h_{i}\bar{z}^{(i)}$. Other sites that do not exceed the threshold, $th$, remain as \(\hat{\sigma}_z^{(i)}\) or \(\hat{\sigma}_z^{(j)}\). If such sites are adjacent to the frozen sites, the interaction term becomes the fourth term. Consequently, the first and the second terms denote the Hamiltonian corresponding to the annealing process. The final SQF Hamiltonian can be rewritten as in Eq.~(\ref{eq:H_ising_SQF_2}), and the singly biased frozen sites contribute to the offset term.

For example, Fig.~\ref{fig:SQF process} (b) shows a histogram representing the energies of samples obtained after each iteration of SQF. SQF was applied to solve the NAE3SAT problem with 150 variables and a 2.1 clause-to-variable ratio. One thousand shots were taken for each iteration with annealing time set to 20\(\mu\)s, and the SQF threshold was set to 0.6. For the following experiments, shot numbers and annealing times were the same if not specified. The result indicates that SQF lowers the average energy of samples and finds some solutions with lower energy that were not collected with the initial annealing (iteration 0). 
The lowest energies were collected after iterations of SQF (iteration 3). This result indicates that SQF enhances solution qualities. We hypothesize that reducing the number of qubits via SQF enhances the energy gap among ground states, thereby reducing the chance of being in the excited states, as shown in the next section.

\begin{figure*}[t]
\includegraphics[width=0.9\textwidth]
{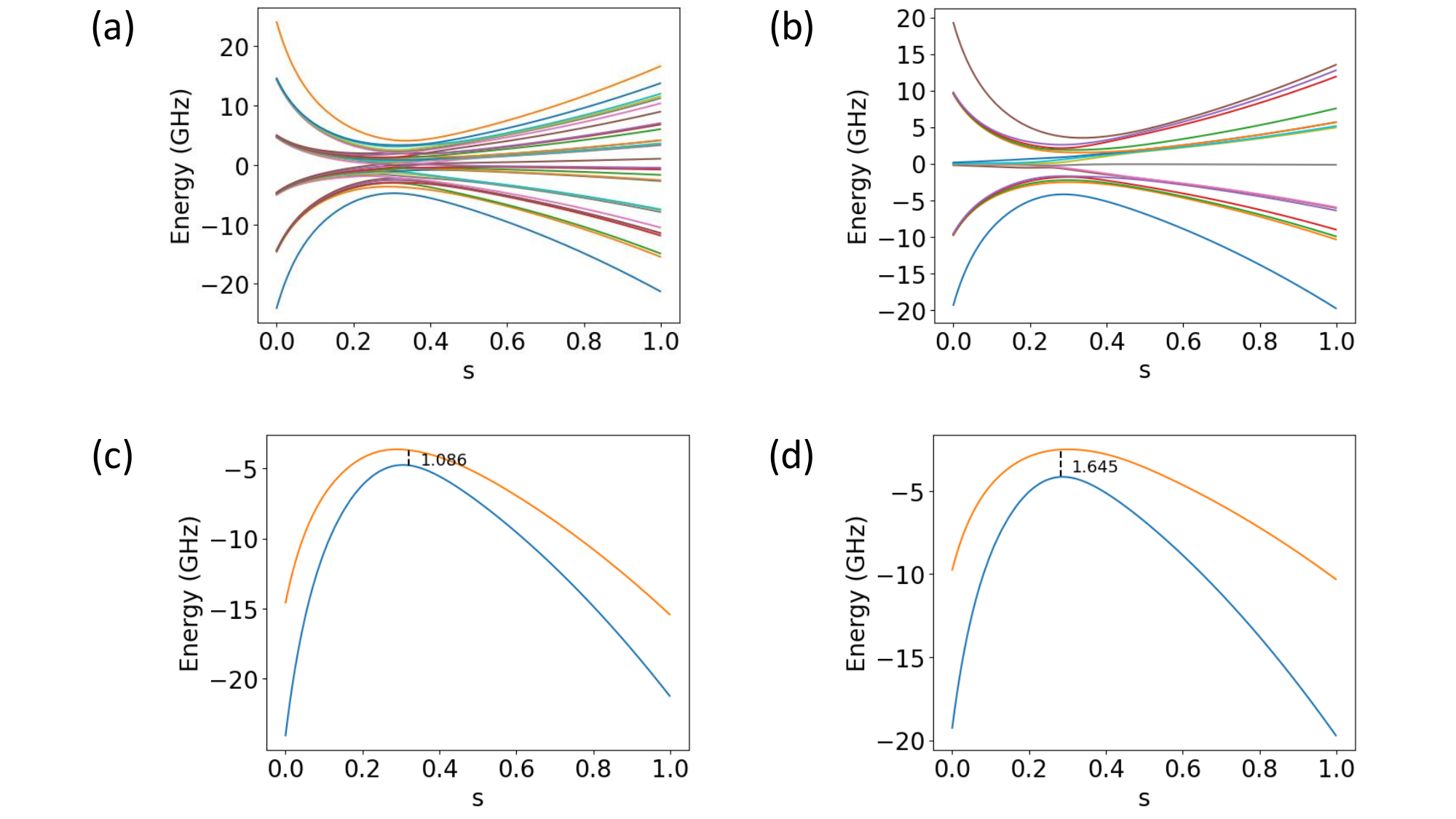}
\caption{\label{fig:energy_gap} Energy gap comparison in the instantaneous energy spectra as a function of anneal fraction $s$ (a) before and (b) after freezing a qubit to a constant value. To demonstrate the effect of SQF on the energy gap separation, we use five qubit random Ising systems with all-to-all connectivity as shown in Fig.~\ref{fig:system} (a). We zoom in on the ground state and the first excited states of the energy spectra without, (c), and with, (d), the SQF process. After freezing, the minimum spectral gap increases from 1.086 GHz to 1.645 GHz (black dashed lines).
}
\end{figure*}

\section{Results: SQF effects with enhanced energy gap} 
Fig.~\ref{fig:energy_gap} describes energy gap enhancement after freezing a qubit in a simple analytical model of Fig.~\ref{fig:system}(a). The test model is a random Ising instance with all-to-all connectivity in which biases and coupling strengths are randomly selected from uniform distributions of ranges from -2 to 2 and -1 to 1, respectively. Figs.~\ref{fig:energy_gap}(a) and (b) are instantaneous eigenenergy spectra during the adiabatic annealing process of Hamiltonians calculated using \textit{Ref:~scipy.linalg.eigvalsh}. The two graphs of Figs.~\ref{fig:energy_gap}(c) and (d) show the difference of the energy spectra of the ground states and the first excited states before and after freezing a value. 

The problem is first solved with a brute force technique, and then the result is analyzed to see which qubit to freeze. As an extreme and ideal case, to show the enhanced spectral gap, the logical qubit with different values for the ground state and the first excited state is frozen to match the value of the ground state. The black line indicates the spectral gap of the system has increased from 1.086 GHz to 1.645 GHz. 

\begin{figure*}[t!]
\includegraphics[width=0.9\textwidth]
{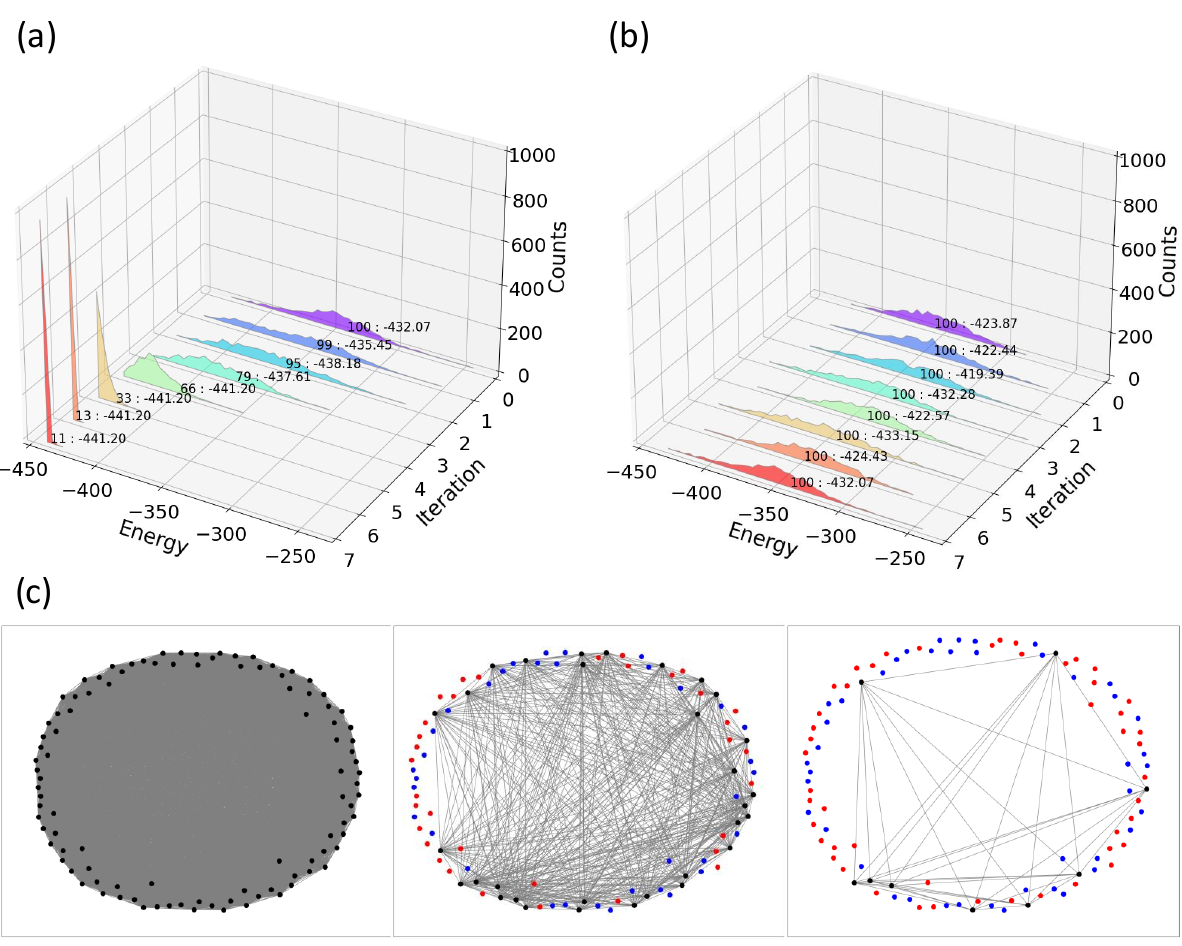}
\caption{\label{fig:random_ising} Application of SQF on random Ising instance for the number of variables $N=$100 (a) with SQF, and (b) without SQF. 
We perform eight iterations for both cases. The number label on each histogram denotes "the number of active qubit variables: the lowest energy found in a run of QA," respectively. The number of samples per iteration is 1000 and the SQF threshold is set to 0.75. With no SQF application, the lowest energy found with 8000 samples is -443.44, while the SQF scheme gradually reduces the number of variables to find the lowest energy -445.81.
The change of the graph of the problem is shown in (c) during SQF iterations. Black circles represent the active variables, while red and blue circles the variables frozen to 1 and -1, respectively. 
}
\end{figure*}

In this test case, we explicitly targeted the qubit to freeze most effectively, as the brute force technique provided the full information of the system. However, in larger systems, we only have access to partial information about the system. To investigate further the effect on larger problems, another random Ising problem with 100 variables that are directly mapped to qubits is created. Fig.~\ref{fig:random_ising}(a) shows histograms of sample energies founded with the initial attempt of quantum annealing (iteration 0) followed by 7 iterations of SQF applications (iteration 1 through 7). Throughout the experiment, the number of shots is set to 1000 and the annealing time to 20\({\mu}s\), SQF threshold 0.7. Upon retrieving sample sets from D-Wave quantum annealer, energy value corresponding to each sample can be converted from arbitrary unit to physical dimension and vice versa with the relation described by the anneal schedule~\cite{dwaveschedule}. 

The lowest energy found with the pure quantum annealing is -432.07 and the number of variables is 100. Successive quantum annealing cycles with the SQF scheme are observed to produce lower energy states. After 7 iterations, the problem size is reduced to 11 variables, and the lowest energy appears to be converged at -441.20. 
For comparison, the same Ising problem is repeatedly executed 8 times with only the original quantum annealer scheme with the same condition to see the statistical behavior of each execution as shown in Fig.~\ref{fig:random_ising}(b). In this case, the lowest energy found was -432.07 and the mean of the lowest energies found was -426.26. With these results, we are able to verify that SQF indeed allows D-Wave quantum annealer to produce a lower energy state. This comparison also suggests that the number of shots can be substantially reduced with the SQF application.

Fig.~\ref{fig:random_ising}(c) shows how the problem graph changes while producing the result in Fig.~\ref{fig:random_ising}(a). Red circles in the graph are the variables frozen to 1, blue circles to -1, black circles are active variables, and solid lines represent connectivity between the variables. This representation visualizes how the problem graph is simplified and the variables are reduced as the SQF scheme is applied.


In order to demonstrate if the SQF scheme can introduce an advantage in finding solutions for a meaningful problem, we can challenge a large-scale NAE3SAT problem with a quantum annealer and SQF scheme. A 3-Satisfiability (3SAT) problem belongs to the class of NP-complete problems that  determines the satisfiability among plural clauses in conjunction. Each clause consists of 3 literals that can be either positive or negative of a boolean variable. NAE3SAT problem is a variant of 3SAT problem where the three variables in each clause are required to be not all equal. The problem was generated using dimod.generator.random\_nae3sat API \cite{dwavenae3sat} provided in D-Wave Ocean SDK. The document states that the problem is formulated so that each clause contributes -1 when satisfied and +3 when violated. Thus, the minimum energy, or the energy of the solution, is lower-bounded by the number of clauses consisting the problem, \(E_\mathrm{min}=-N_\mathrm{cl}\), and the satisfaction ratio can be calculated, \(R_\mathrm{sat}=1-(H(s)-E_\mathrm{min})/(4 N_\mathrm{cl})\), where $H(s)$ is the sample energy and $N_\mathrm{cl}$ is the number of clauses.

\begin{figure*}[htb]
\includegraphics[width=0.9\textwidth]{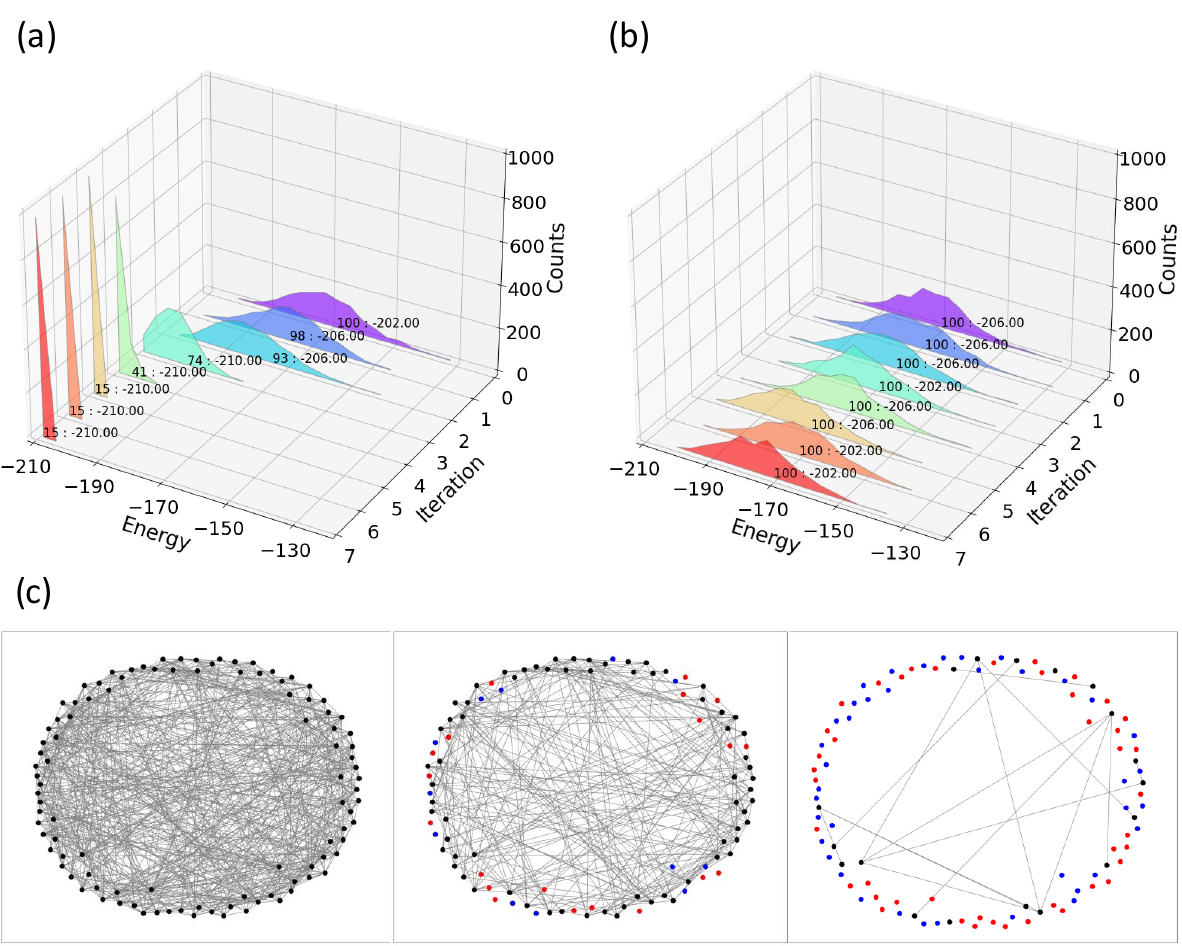}
\caption{\label{fig:nae3sat} Application of SQF on NAE3SAT for the number of variables $N=$100 (a) with SQF, and (b) without SQF. 
We perform eight iterations for both cases. The number label on each histogram denotes ``the number of active qubit variables: the lowest energy found in a run of QA,'' respectively. The number of samples per iteration is 1000 and the SQF threshold is set to 0.75. With no SQF application, the lowest energy found with 8000 samples is -206, while the SQF scheme gradually reduces the number of variables to find the lowest energy -210, which is the ground state energy of the system.  
The change of the graph of the problem  is shown in (c) during SQF iterations. Black circles represent the active variables, while red and blue circles the variables frozen to 1 and -1, respectively. 
}
\end{figure*}

Figure~\ref{fig:nae3sat}(b) shows the energy histogram evolution while solving the NAE3SAT problem consisting of 100 variables with clause-to-variable ratio of 2.1 by applying the SQF scheme. For the following experiments with the NAE3SAT problem, the clause-to-variable ratio was set to 2.1 regardless of the problem size. Similarly to the previous example, the problem size is observed to be reduced, and the lowest energies decrease from -206 to -210 with SQF iterations; \(R_\mathrm{sat}\) is increased from 0.99 to 1. 
That is to say, after applying SQF, two more clauses are satisfied, with which we obtain the ground state of the system. The result indicates that SQF seems to make a quantum annealer scalable and effective in larger problems in comparison with the results of  conventional D-Wave quantum annealing without SQF as shown in Fig.~\ref{fig:nae3sat}(b). The reference histogram is obtained by solving the same NAE3SAT problem 8 times with a shot number of 1000. The improved satisfaction ratio is only 0.01 higher than the original result, but we note that the ground state solution unreachable with the traditional QA is found by applying SQF.

\section{Searching for a global minimum as the ground state}

\begin{figure*}[htb]
\includegraphics[width=1.0\textwidth]
{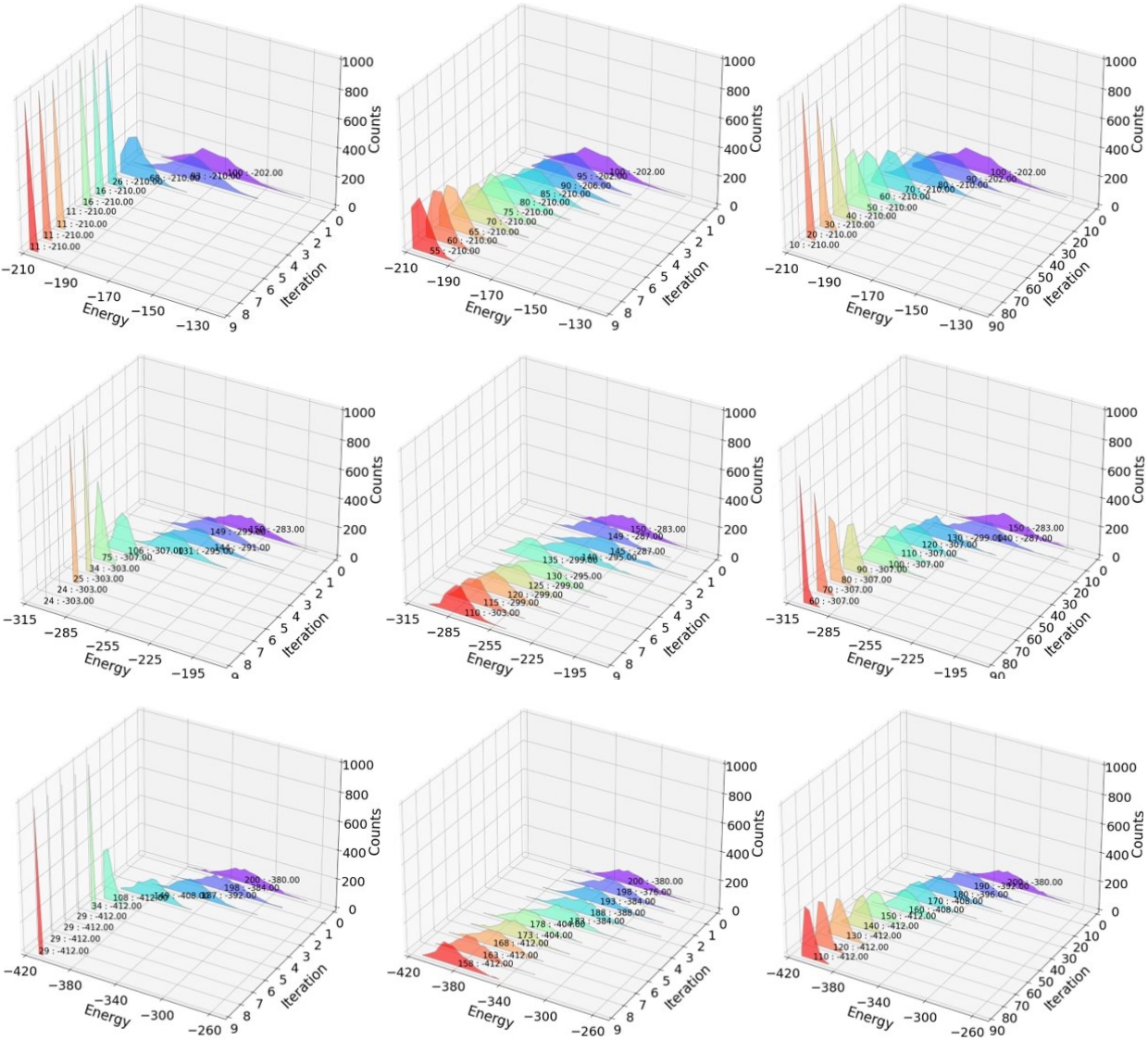}
\caption{\label{fig:all_strategies} Result of three different SQF strategies tested on NAE3SAT problems. Each column, sequentially, represents one of the three strategies: ``progressive threshold'', ``first-M'', and ``one each time”. Each row, from top to bottom, corresponds to a different problem size of \(N = 100, 150, 200\). For strategies of ``progressive threshold'' and ``first-M'', SQF iteration was applied 9 times, and for the strategy of ``one each time'', the iteration was applied 90 times so that 90 variables were frozen in total.}
\end{figure*}

Quantum annealing does not guarantee to find a global minimum when adiabatic conditions are not satisfied. However, we can investigate how close the SQF scheme approaches to a global minimum. 
In addition to the vanilla SQF scheme explained above, we tested three different strategies for freezing schemes to see if the SQF performance improves. With the first strategy of ``progressive threshold,'' we increase the freezing threshold gradually as we freeze qubits because the reduced problem size will likely result in a more polarized statistical distribution. Here, there is no limit against the number of qubits to freeze. 
The second strategy of ``first-M''is to freeze only the first $M$ qubits in descending order of analyzed tendencies. 
Rather than freezing many qubits at once, we can anticipate to improve the performance by avoiding being captured in local minima if we freeze qubits more conservatively. The last strategy of ``one each time'' is to freeze only one qubit with the highest likeliness in each iteration of SQF without calculating the energy impact of freezing. Such a strategy will cause higher time complexity but only linearly proportional to the number of qubits. 

In order to verify if the QA result corresponds to a global minimum, we can try these strategies with NAE3SAT problems with three different problem sizes of N = 100, 150, and 200. Similarly to the previous experiments, the number of shots and annealing times are set to 1000 and 20 \(\mu\)s, respectively, and the SQF threshold is set to 0.6.

Fig.~\ref{fig:all_strategies} shows the results of three strategies on the various problem sizes. Each column, from left to right, corresponds to one of the three strategies. Each row, from top to bottom, corresponds to a different problem size. Like the previous energy histograms, the number label on each histogram indicates the remaining number of variables, and the lowest energy.

With the first strategy of ``progressive threshold,'' the SQF threshold is increased by 0.05 after every three iterations of SQF. We can find the ground state of the problem \(N = 100\) along with the increased satisfaction ratio from 0.99 to 1, but for larger problems, SQF results in better quality solutions while seeming to fall in local minima. In terms of satisfaction ratio, for \(N = 150\), the ratio increases from 0.975 to 0.990, and for \(N = 200\), from 0.976 to 0.995.  
With the second strategy of ``fisrt-M,'' we freeze only up to 5 qubits in each iteration of SQF, and the ground state of the smallest problem, \(N = 100\), is found as well. However, likewise, for larger problems, only sub-optimal states are found. In addition, more SQF iterations are needed to reach solutions with the same quality as strategy ``progressive threshold''. 
The last strategy shows a similar trend. The ground state of the smallest problem can be found, but not for larger problem sizes. Overall, the results indicate that all the strategies are equally effective as their lowest energy values are the same; the first strategy ``progressive threshold''seems to be the most efficient as it reached the ground state or the sub-optimal local minima with the least SQF iterations.
In summary, while each is capable of reaching the same lowest energy state, the strategy of ``progressive threshold'' achieves the fewest iterations of SQF to capture the solution state. 

\section{Discussion}
We introduced a heuristic scheme called statistical qubit freezing which we tested across various problems and strategies to overcome the physical limitations of quantum annealing device scalability. 
Our results demonstrate that SQF significantly enhances the capability of QA devices to identify the ground state in large-scale problems that are typically unresolvable through conventional quantum annealing alone.
In tests involving larger problem scales with variable counts of \(N = 150\) and \(200\), SQF enabled QA devices to achieve lower energy states. In these experiments, we closely examined the influence of key SQF parameters, including the SQF threshold and the maximum number of qubits frozen in each iteration of SQF. 

The experimental results showed that while various strategies achieved similar effectiveness, meaning that each may reach the same lowest energy state by running the QA multiple times, the strategy of progressively increasing the SQF threshold without limiting the number of qubits frozen proved most efficient. This approach required the fewest iterations of SQF among the three different strategies tested to reach the solution states with the lowest energies. However, there was noticeable occasions for the QA process trapped in local minima. Nevertheless, the verification of solution quality before freezing variables ensures that the SQF scheme has a positive impact overall. Compared to the previous works, SQF with such additional verification steps allows quantum annealing devices to find lower energy solutions in a more reliable and accelerated manner.

The D-Wave quantum annealer operates at a temperature below 20 mK \cite{dwaveimplementation} with a fixed energy scale range \cite{dwaveschedule}. Given this fixed energy scale, as more qubits are involved in a problem, the eigenenergy spectrum of the system becomes more densely populated. If the spectral gap of the system becomes comparable to or smaller than the thermal noise, the system inevitably excites during the annealing process. By converting arbitrary energy values to physical energy values using the annealing function \cite{dwaveschedule}, we determined that the energy scale of the system discussed in this paper is approximately $\sim10^3 \mathrm{GHz}$. Assuming an operating temperature of 20 mK, equivalent to 0.4167 GHz, a straightforward calculation suggests that systems with a few thousand eigenstates are already too dense to remain in the ground state during the annealing process. In this context, SQF appears to effectively widen the energy gaps of energy states. As demonstrated by the experimental results shown in the first row, third column of Figure \ref{fig:all_strategies}, for the NAE3SAT problem with N = 100, the sampled energies are focused on the ground state when the problem size is reduced to 40 variables or fewer.

Compared to other error-correcting or error-mitigating methodologies and post-processing techniques, SQF operates independently of any prior knowledge regarding error models, device architecture, or the need for redundant qubit usage. Our experimental results indicate that by merely collecting samples and accordingly reducing the size of the problem Hamiltonian based on sample statistics, the QA device is able to locate a state of lower energy.

Additionally, SQF demonstrates advantage in facilitating the minor-embedding process, which is an essential step to run a typical QA device. As problem sizes escalate under limitations in qubit connectivity, long chain lengths to form logical qubits with the sequence of physical qubits typically lead to a degradation in solution accuracy. In addition, the computational time required to determine a suitable minor embedding also increases. However, iterative reduction of the problem size by SQF effectively shortens the chain length and reduces the time required for minor embedding in subsequent annealing cycles. SQF can eventually enhance the physical limit of scalability of the natural QA significantly.

For future research, it is essential to devise a more systematic SQF process and derive the relationship between SQF governing variables and solution quality. This will enable us to identify the preconditions necessary for consistently reaching the global minima. Furthermore, developing a methodology to effectively escape local minima is critical for achieving the ultimate performance objectives of SQF. Finally, a formal mathematical analysis will provide a robust theoretical foundation, to uncover the efficacy and potential limitations of the SQF algorithm. \\

\iftrue
\begin{acknowledgments}
We acknowledge the support of LG Uplus and the National Research Foundation of Korea (NRF) grant funded by the Korean government (MSIT) (No. 2021R1A2C2013790).
\end{acknowledgments}
\fi
\nocite{*}
\bibliography{apssamp}

\end{document}